**Social interaction networks and depressive symptoms**

Timon Elmer[1] & Christoph Stadtfeld[1]

[1]Chair of Social Networks

ETH Zürich

Department of Humanities, Social and Political Sciences

Weinbergstrasse 109, 8092 Zürich, Switzerland

Abstract

Face-to-face social interactions are an important aspect of peoples' social lives. A lack of interactions can explain how individuals develop depressive symptoms, but depressive symptoms can also explain how individuals engage in social interactions. Understanding in detail how depression affects individuals' social interaction networks is important to break this vicious cycle of social isolation and depression. This article tackles two central methodological challenges in understanding the micro-level mechanisms between depressive symptoms and social interactions. The first contribution is the application of novel data collection strategies that employ RFID sensors to record behavioral data on interpersonal interaction, which we combine with self-reported depressive symptoms and sociometric data on friendship relations. The second contribution is the analysis of these data with statistical social network methodology that allows testing hypotheses on the duration of interactions between pairs of individuals. With this unique approach, we test four social network hypotheses in an empirical setting of two first-year undergraduate student cohorts ( $N_{pairs} = 2,454$, $N_{individuals} = 123$) spending a weekend together in a remote camp house. We conclude that depressive symptoms are associated with (1) spending less time in social interaction, (2) spending time with similarly depressed others, (3) spending time in pair-wise interactions rather than group interactions, and with (4) spending more time with reciprocal friends (but not with unilaterally perceived friends). Our findings offer new insights into social consequences of depressive symptoms and call for the development of social network-oriented intervention strategies to prevent depressed individuals from being socially isolated.

## Introduction

Social interactions are the smallest building blocks of interpersonal social network and are a prerequisite of the formation of functional social relationships. A lack of social interactions has detrimental effects on an individual's physical and psychological health. On the physical side, socially isolated individuals have a higher rate of coronary heart disease, stroke, and mortality (Holt-Lunstad, Smith, Baker, Harris, & Stephenson, 2015; Steptoe, Shankar, Demakakos, & Wardle, 2013; Valtorta, Kanaan, Gilbody, Ronzi, & Hanratty, 2016). But social isolation also has a strong effect on psychological health such as depressive symptoms (Jose & Lim, 2014; Kawachi & Berkman, 2001).

Social isolation and depression are tightly linked; not only does social isolation lead to more depressive symptoms but also how an individual integrates into a community is dependent on the person's degree of depressive symptoms (e.g., Elmer, Boda, & Stadtfeld, 2017; Schaefer, Kornienko, & Fox, 2011). This lack of social integration has been linked to the functional and social impairment associated with clinical and sub-clinical levels of depressive symptoms (Backenstrass et al., 2006; Gotlib, Lewinsohn, & Seeley, 1995). It has been argued further that the social relationships of depressed individuals show increased levels of interpersonal distress and dysfunction (Barnett & Gotlib, 1988; Segrin, 2000). Depressed individuals are thus less likely to have functional social relationships, which can further increase their depressive symptoms, leading to a vicious cycle of social isolation and depression.

Social impairment related to depression is expected to manifest in social interactions, and thus shape how they evolve. Although much research has been conducted on the social nature of depression, relatively little is known on how depressive symptoms are associated with individual's behavior patterns in social interaction networks.

In this study we focus on this fine-grained type of social relations and ask how social interactions are affected by depressive symptoms, and how preexisting friendship ties moderate this relation. We situate our study in a context where individuals (first-week undergraduate students) get to know each other in the process of an emerging social group. Rather than using self-reports of social interactions, we collected fine-grained data on face-to-face interactions using newly developed and validated Radio Frequency Identification Devices (RFID; Cattuto et al., 2010; Elmer, Chaitanya, Purwar, & Stadtfeld, 2018). Those automatically record when study participants face each other in very close proximity that is typically associated with a social interaction. These data are combined with self-reported data on friendship relations and depressive symptoms assessed prior to the social interactions. We apply state of the art statistical social network analysis methods (Dekker, Krackhardt, & Snijders, 2007; Krackhardt, 1988) that take into account that relational observations are not statistically independent. We, thereby, test four preregistered[1] hypotheses on the interplay between social interactions, friendship, and depressive symptoms. The unique social networks design allows us to test these relation hypotheses that require data of a closed group of interacting individuals and data on the depressive symptoms of (possibly) all individuals in the group.

The first hypothesis (depression-isolation hypothesis) states that depressive symptoms are associated with spending less time in social interactions with others. These differences might be related to a lack of social skills associated with depressive symptoms (Lewinsohn, 1974). This can affect the quantity of social interactions of depressed individuals in potentially two ways: (1)

---

[1] www.osf.io/xce9g

Individuals with depressive symptoms elicit rejection from others as they induce a negative mood in their interaction partners (as stated in Coyne's interactional theory of depression; Coyne, 1976a, 1976b) and (2) depressed individuals receive less reinforcement from the social environment which contributes to a feeling of discomfort in social interactions and decreased social participation (Brown, Strauman, Barrantes-Vidal, Silvia, & Kwapil, 2011; Libet & Lewinsohn, 1973; Segrin, 2000). In prior work, Brown and colleagues (2011) have reported a negative association between depressive symptoms and the amount of self-reported social interactions. At the same time, other studies reported no differences in the quantity of social interactions but only on qualitative aspects of social interactions (Baddeley, Pennebaker, & Beevers, 2012; Nezlek, Hampton, & Shean, 2000; Nezlek, Imbrie, & Shean, 1994). The use of a direct behavioral measure of social interactions allows us to overcome measurement biases that are associated with how depressed individuals report social interactions (e.g., having more negative social (self-)perceptions; Gadassi & Rafaeli, 2015; Gotlib, 1983).

The second hypothesis (depression-homophily hypothesis) states that individuals will be more likely to interact with others who have a similar level of depression. This hypothesis was first developed in the context of friendship networks (e.g., Elmer et al., 2017; Schaefer et al., 2011). The tendency to bond with similar others (homophily; McPherson, Smith-Lovin, & Cook, 2001) has been found to be one of the most consistent patterns in social networks. It is expected to be prevalent on the depression scale, as sharing emotional states with similar others can lead to more compassion and self-disclosure and thus to more rewarding interactions (Rook, Pietromonaco, & Lewis, 1994).

The third hypothesis (depression-friendship hypothesis) proposes that individuals' depressive symptoms are associated with the extent to which they interact with friends. The direction of this association, however, is unclear. We assume that friends tend to spend time together (Friedkin, 1990) and that they will be more aware of each others' mental health than non-friends (e.g., through signs of verbal or non-verbal behaviors in previous interactions; Segrin, 2000). The literature on the association between friendship and interaction proposes different arguments. On the one hand, it is argued that individuals with more depressive symptoms tend to spend more time with friends talking about (negative) feelings (Berry & Hansen, 1996; Rose, 2002). On the other hand, Coyne (1976b) proposes that interactions with depressed individuals are avoided because such interactions induce a negative affect in others. Empirical support for latter mechanism is provided by Brown et al. (2011), who showed that depressed individuals tend to interact relatively less with others whom they perceive as close (e.g., friends).

The fourth hypothesis (dyadic-isolation hypothesis) states that individuals with depressive symptoms will be more likely to interact within pairs (dyads), rather than interacting in groups of three or more. We assume that dyadic interactions are more frequent than group interactions for depressed individuals because of depressed individuals' tendency of "discussing and revisiting problems, speculating about problems, and focusing on negative feelings" (i.e., co-rumination; Rose, 2002, p. 1830). If co-rumination is more likely to occur in pairs, this could lead to an over-representation of dyadic interactions among depressed individuals.

The present study was designed to remedy an empirical gap by testing four hypotheses that relate to interaction patterns of individuals with different levels of depression. We test these with social network methods in a setting of automatically collected data on social interactions. Understanding the interpersonal consequences of depressive symptoms at such a fine-grained level can guide the design of future interventions to break the vicious cycle of social isolation and depressive symptoms.

## Methods

### Participants

We investigated our research questions with two independent datasets newly formed undergraduate student cohorts attending a voluntary social event on the first (sample one) and second (sample two) weekend of their studies. The data was collected in the context of the *Swiss StudentLife* study (Stadtfeld, Vörös, Elmer, Boda, & Raabe, in press). The first sample consisted of $N_1 = 73$ individuals, of which 14 individuals belonged to the student organization that organized the event. The second sample consisted of $N_2 = 50$ individuals, including 14 student organization members. Prior to the weekend, 53 (73%; Sample one) and 48 (96%; Sample two) of the participants administered an online survey that assessed friendship ties within the cohort and depressive symptoms. None of the student organization members of sample one participated in the survey. All non-responses were treated as missing data. The first sample was predominately male (37% female) whereas the second sample was mostly female (60%). The mean age of the two samples were 20.75 years ($SD = 2.09$) and 21.73 years ($SD = 3.24$), respectively. In total, there were 3,853 dyadic relations ($N_1^{dyads} = \frac{N_1(N_1-1)}{2} = 2,628$; $N_2^{dyads} = 1,225$) of which 2,454 (64%) remain after listwise deletion of missing data. Hence, the sample size for our analyses should be sufficiently large.

**Procedure**

In the three days prior to the weekend (Tuesday to Thursday) participants were invited to administer the online questionnaire. The study was advertised as a broad investigation about social integration and the lives of students in the first year at university.

Before the arrival at the remotely located camp house, each participant was equipped with a badge that consisted of the active Radio Frequency Identification device (RFID), which allowed us to measure their social interactions (Cattuto et al., 2010; Elmer et al., 2018). The badge was covered with a piece of paper with the participants' name printed on it. Hence, the RFID badge was not visible. Participants were briefed on the badge's functionality and purpose of application. All participants were instructed to wear the RFID badge during their time spent awake and place them on chest height. In both samples all of the participants agreed to wearing the badge throughout the weekend. During the event, study confederates checked that the participants wore the badge correctly. The event was scheduled from Friday 7pm to Sunday 8pm (sample one) and from Saturday 3pm to Sunday 11pm (sample two). During the course of the weekend there were some organized activities (e.g., group games, lectures), but most of the time was unstructured so that participants could freely interact with each other.

**Measures**

**Social Interactions.**

During the course of the weekend social interactions were assessed using active Radio Frequency Identification (RFID) badges. The hardware was constituted of 2.4 GHz RFID badges with realtime proximity and position tracking utilizing the Bluetooth low energy protocol.

To detect the signal between two RFID badges, both badges need to be close to each other (range 1-1.5m; Cattuto et al. 2010) and to an RFID reader. RFID readers are designed to receive signals from RFID badges that are in the range of 10 meters from a reader. Before arrival of the participants, the camp house was equipped with 8 RFID readers so that in every room of the house and in commonly used outside areas (e.g., smoking area) signals between RFID badges could be detected. We followed the recommendations by Elmer et al. (2018), to enhance the validity of RFID badges by merging interactions of the same dyad if the signals are no longer than 75 seconds apart. Robustness analyses conducted on data that was not processed in that way can be found in Table S1 and Table S2 of the Supplementary Materials. More details on the RFID badges to measure face-to-face interactions can be found elsewhere (Cattuto et al., 2010; Elmer et al., 2018).

Existing studies predominately use self-report measures to assess social interactions. Biases in self-reports of individuals with depressive symptoms might contribute to differences in their self-reported interactions, as – for instance – depressed individuals tend to view things more negatively than non-depressed (Gotlib, 1983). With the RFID based method of social interaction measurement, we aim to overcome these biases.

**Friendship ties.**

Friendship ties were measured with the items "which of your fellow students would you call friends?" (German original: "Welche Deiner Mitstudierenden würdest Du als Freunde bezeichnen?"). Below the item were 20 name generators displayed (i.e., text boxes where participants could enter the names of the individuals). An auto-complete function suggested the full names of other participants when starting to type in this text field. The nominations of that item were used to construct a binary adjacency matrix $A$ where each entry $a_{ij}$ represents the nomination of individual $j$ by individual $i$ (0 = no nomination, 1 = nomination).

Because our statistical method requires the independent variables to be symmetric matrices (for details see Section Statistical Analyses), we constructed a symmetrized friendship matrix indicating if at least one $i \rightarrow j$ or $j \rightarrow i$ friendship nomination was present. To explore the unique contribution of weak and strong friendship ties, two additional adjacency matrices were created in which cells indicate if the tie is (i) a mutual (strong) friendship tie (i.e., $i \rightarrow j$ and $j \rightarrow i$) or (ii) a asymmetric (weak) friendship tie (i.e., either $i \rightarrow j$ or $j \rightarrow i$, but not a mutual tie). These measures are used for explorations of friendship strength (Friedkin, 1990).

### Depressive Symptoms.

Depressive symptoms were measured with the German version of the Center for Epidemiologic Studies Depression Scale – Revised (Hautzinger & Bailer, 1993) with 20 items on a 4-point scale ranging from 0 (occurred never or rarely) to 3 (occurred most of the time or always) reflecting how often the respective symptoms was experienced during the preceding week. Sample items are for instance "feeling depressed" or "feeling everything one does is an effort". The depression score was computed by taking the sum of all 20 items. The items of this scale were highly internally consistent (Cronbach's alpha = .84).

## Statistical Analyses

We investigate our research questions using Multiple Regression Quadratic Assignment Procedures (MRQAP; Dekker et al., 2007; Krackhardt, 1988). In social network analysis MRQAPs are considered a core method to analyze weighted networks. Mathematically, a

MRQAP is defined similarly to a linear regression model but with data arranged in matrices instead of vectors:

$$y_{ij} = \beta_0 + \sum_{k=1}^{m} \beta_k \left( x_{ij}^k \right) + e_{ij}$$

where $y$ is the dependent matrix and $m$ is the number of independent matrices $x^k$. Parameters $\beta_k$ are coefficients and $e_{ij}$ the error terms.

Indexes $i$ and $j$ represent two individuals in a given matrix. If an $x^k$ represents a friendship network, $x_{ij}^k$ would indicate that $i$ considers $j$ a friend. Similarly, an $x^k$ could represent the similarity between individuals with $x_{ij}^k$, for example, indicating the difference in depressive symptoms of individuals $i$ and $j$ (i.e., depression similarity). In principle, parameters of a MRQAPs can be interpreted like parameters of a linear regression model, as they are estimated with ordinal least squares (OLS) estimators. MRQAPs differ only in two ways from linear regression models. The first difference is that, the unit of analysis in MRQAPs is on the dyadic level. Hence, the dependent variable is an adjacency matrix of dyadic relations (i.e., $y_{ij}$ is the time individuals $i$ and $j$ interacted). Also, the independent variables of a MRQAP need to be defined on a dyadic level. Examples for friendship and depression similarity are given above. The second difference to linear regression models is concerned with the independence assumption. Social network data violate the assumption of independent observations: For instance, a person A's interactions with Person B cannot be assumed to be independent of Person A's interactions with Person C. Because characteristics of Person A - e.g., being female - affect both interactions. For this reason, the standard errors obtained through OLS estimation cannot be used for statistical inference. MRQAPs consider the dependencies between observations in the estimation of standard errors by relying on permutation tests for statistical inference: The OLS regression results obtained with the observed adjacency matrix are compared to a large number of regression results in which the dependent matrix $y$ has been permuted. According to Dekker,

Krackhardt, and Snijders (2007), the Y-permuted MRQAPs are (among the MRQAP methods)

the most conservative method to obtain statistical inference—others are for example

permutations of the independent variables. When permuting the dependent matrix $y$, random

rows and columns are swapped, while the independent variables $x$ remain unaffected. This way

structural aspects of the dependent network are preserved (e.g., the outdegree distribution), while

generating a distribution that assumes no association between $y$ and the $x^k$. Because of the

permutation based statistical inference of the MRQAP framework, no standard errors or

confidence intervals of the estimates can be computed. Thus, we rely on p-values for statistical

inference. However, we also report the results of a multivariate linear regression model in Table

S1 of the Supplementary Materials, where confidence intervals are reported. Within the MRQAP

framework, the p-value is calculated based on the percent rank of the estimate of the observed network in the distribution of estimates based on permuted networks. For instance, the percent rank of .99 indicates that 99 percent of the coefficient based on permuted networks are smaller or equal to the observed estimate. The probability of observing larger estimates under the null-hypothesis is thus p = .01 (for details see Dekker et al., 2007; Krackhardt, 1988).

 We analyze the two samples jointly and, therefore, use a *multi-group* MRQAP, in which the dependent matrices $y$ of the two samples are permuted separately (Burnett Heyes et al., 2015;

Dekker et al., 2007).

 The dependent matrix in our study was constructed by summing the duration of each social interaction of each dyad. Hence, the adjacency matrix is undirected, symmetric, and weighted. Because the distribution of the residuals of the MRQAP model with this dependent matrix was highly skewed (s = 4.00), the linear regression assumption of normality of errors was violated. Thus, we log-transformed the dependent matrix (skewness of residuals after transformation: s = 0.26) following standard procedure in linear regression models. In Table S1 of the Supplementary Materials we also report results based on non-transformed variables.

The independent matrices $x_{ij}$ in our MRQAP model represent either dyad-level aggregates of individual's attributes (e.g., the difference in age of the two individuals) or dyadic relations (e.g., friendship nominations).

We test *depression isolation* with the *depression mean* matrix, where each entry constitutes the mean depression score of both individuals *i* and *j*. *Depression homophily* is tested with the *depression similarity* matrix, which consists of values representing the degree of similarity in depression ($x_{ij} = (-1)|v_i - v_j|$, were $v_i$ is the depression value for individual *i*). Given that the reference category for the depression similarity effect is being identical on the depression score, the "raw" depression mean effect can be interpreted as the effect of both individuals being equally depressed. We included an interaction of these two matrices to account for differences in the importance of homophilic processes depending on the levels of depression.

To what extent depressed individuals interact with their friends is tested with interactions of the depression mean and a friendship matrix. Friendship was defined when at least one of two individuals of a dyad reported a friendship tie. Whether or not friendship is mutual or asymmetric can be potentially relevant and serve as an indicator of relationship strength (Friedkin, 1990). For this reason, we conducted additional analyses in which we also consider the mutual and asymmetric friendship ties as separate independent matrices (i.e., a binary matrix indicating when both individuals nominated each other as friends and a binary matrix indicating whether or not exactly one individual of the dyad nominated the other as a friend).

Additionally, we included a dummy variable indicating whether or not the data was collected in sample two. To control for the effect of gender, we added dummy matrices as independent variables for the case of at least one female being in the interaction and for both individuals being female. Age-related effects were included with a centered age mean matrix ($x_{ij}^k = \frac{(v_i - \overline{v}) + (v_j - \overline{v})}{2}$) and an age similarity matrix ($x_{ij}^k = (-1)|v_i - v_j|$, where $v_i$ is the age value for individual $i$).

*Dyadic isolation* is evaluated outside the MRQAP framework. For this, we computed the number of seconds that each individual spent in either a dyadic or group interaction (i.e., at least three individuals present in the social interaction). These two variables are then compared to each other with respect to an individual's depression scores to assess the degree of dyadic isolation.

## Results

### Description of the data

On average, individuals reported a depression score of 10.28 (*SD* = 5.25) in sample one and 11.98 (*SD* = 7.97) in sample two. According to the screening criteria defined by Radloff (1977), 15 % of the respondents in sample one and 29 % of sample two show clinically relevant levels of depressive symptoms. In sample one, we also collected data of individuals that were in the same study group but chose not attend this voluntary social event or signed up after all slots have been taken. Those individuals attending the weekend did not differ in their level depressive symptoms from those that did not attend this voluntary event (N = 119), t(174) = 0.15, p = .881. A total number of 23,452 social interaction events were recorded in sample one and 12,225 in sample two. These numbers relate to the raw data of recorded RFID interactions over the whole weekend. The average duration of interactions were 94.51 ($SD$ = 212.77) seconds and 86.81 ($SD$ = 186.32) seconds, respectively. The large standard deviation indicates the amount of variability between pairs of students. These social interactions were aggregated to one adjacency matrix per sample where each entry represents the total duration of social interactions between individual $i$ and $j$. Each participant on average interacted 16.87 hours ($SD$ = 7.27) with others in sample one and 11.79 hours ($SD$ = 6.41) in sample two. Figure 2 shows these interaction networks. Each individual is represented as a node, where the node color indicates the degree of depressive symptoms (dark red = high, yellow = low, grey = missing value). The thickness of ties denotes how long two individuals have interacted with each other. The networks exhibit typical social network structures — for example, one can see that interactions tend to cluster within certain regions of the network.

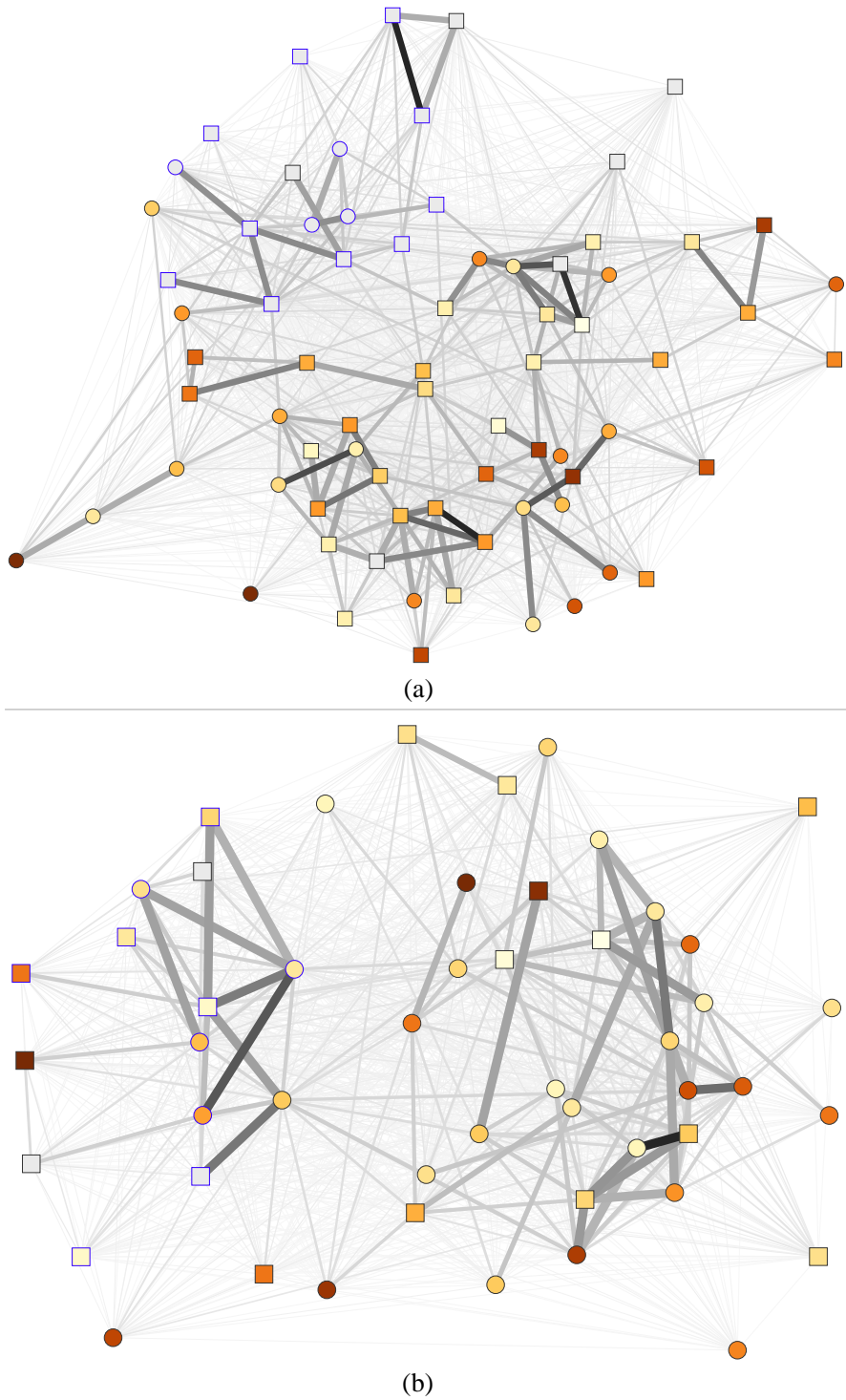

(a)

(b)

*Figure 2.* Durations of social interactions over the course of the data collection for sample one (a) and sample two (b); tie color and width = interaction duration, blue node frame = student organization member, color = depressive symptoms (dark red = high, yellow = low, grey = missing value), circles = females, squares = males, plotted with *visone* (Brandes & Wagner, 2004).

On average, the participants reported 0.66 friendship ties ($SD = 1.28$) in sample one and 2.14 ($SD = 2.13$) in sample two. Because the participants of sample two knew each other for a week longer, more friendship relations were established. In total, 48 ties (sample one) and 107 ties (sample two) were reported. Of those 20 were mutual and 28 were asymmetric in sample one. In sample two, 78 friendship ties were mutual and 29 were asymmetric.

Before testing our hypotheses in a multivariate way on the dyadic level with MRQAPs, we – in the next paragraph – show how depressive symptoms and different aspects of social interactions correlate bivariately on the individual level. Table 1shows the correlation coefficients of depressive symptoms with properties of the interaction network. These coefficients show that depressive symptoms are negatively correlated with how much time individuals spend in social interactions. Depressive symptoms do not correlate with the amount of time spent with friends (symmetrized measure). However, there is a negative correlation with the amount of time spent with mutual friends. We find no evidence for a correlation between depressive symptoms and the amount of time spent in dyadic interactions, but a negative correlation with the amount of time spent in group interactions. These differences of dyadic and group interactions are also reflected in the positive correlation of depressive symptoms with one's ratio of dyadic interactions in all social interactions.

Table 1

*Pearson correlations between depressive symptoms and interaction aggregates*

|  | (1) | (2) | (3) | (4) | (5) | (6) | (7) | (8) | (9) |
|---|---|---|---|---|---|---|---|---|---|
| Depression (1) |  |  |  |  |  |  |  |  |  |
| Age (2) | -.13 |  |  |  |  |  |  |  |  |
| Gender (1 = female)[1] (3) | .17 | -.11 |  |  |  |  |  |  |  |
| T in interaction (4) | -.23* | .08 | .04 |  |  |  |  |  |  |
| T per friend (5) | -.07 | -.11 | -.03 | .44*** |  |  |  |  |  |
| T per mutual friend (6) | -.30* | .09 | .00 | .38** | .91*** |  |  |  |  |
| T per asymmetric friend (7) | -.08 | -.11 | -.05 | .44*** | .99*** | .91*** |  |  |  |
| T in dyadic interactions (8) | -.08 | .12 | .14 | .76*** | .22* | .28* | .22* |  |  |
| T in group interactions (9) | -.26** | .03 | -.01 | .91*** | .46*** | .35** | .46*** | .42*** |  |
| ratio dyadic interactions (10) | .26** | .00 | .07 | -.53*** | -.38*** | -.22 | -.39*** | .09 | -.80*** |

*Note.* $N_{Sample1+2} = 123$. * $p < .05$, ** $p < .01$, *** $p < .001$. T = time [in sec] normalized by the hour (so that the two samples are comparable). [1]Spearman's rank correlations.

**Multi-group MRQAPs**

To test the multivariate relationships between social interactions and individual's attributes, we conducted a multi-group MRQAP analysis (Dekker et al., 2007). The result of this analysis is shown in Table 2, reporting the estimates of observed network $\widehat{\beta}$ and the comparison with the $\beta$ estimated under 5,000 network permutations. The mean value of the estimate under the permuted dependent networks is indicated by $E(\beta)$.

Table 2

*Multi-group QAP results on log transformed interaction durations of dyads*

|  | Estimate | p | E(Est.) | Percentiles | |
|---|---|---|---|---|---|
|  |  |  |  | 2.5th | 97.5th |
| Intercept | 2.504** | .005 | 1.820 | 1.290 | 2.346 |
| Sample two | 0.806 | .344 | 0.835 | 0.693 | 0.974 |
| At least one female | -0.095 | .129 | -0.003 | -0.160 | 0.155 |
| Both female | -0.148* | .036 | 0.004 | -0.158 | 0.160 |
| Age mean (centered) | 0.065* | .013 | 0.000 | -0.059 | 0.057 |
| Age similarity | 0.042** | .009 | 0.000 | -0.035 | 0.035 |
| One student organization | -0.028 | .450 | -0.001 | -0.452 | 0.451 |
| Same student status | 0.269 | .115 | -0.003 | -0.456 | 0.435 |
| Being friends | 2.128*** | <.001 | 0.007 | -0.453 | 0.477 |
| Depression mean | -0.059*** | <.001 | 0.000 | -0.023 | 0.024 |
| Depression similarity | 0.047** | .004 | 0.000 | -0.035 | 0.034 |
| Depression mean * depression similarity | -0.004*** | .001 | 0.000 | -0.002 | 0.002 |
| Depression mean * being friends | -0.012 | .333 | 0.000 | -0.053 | 0.052 |
| $R^2$ | .123 |  |  |  |  |
| Adj. $R^2$ | .119 |  |  |  |  |

*Note.* $N_{sample1+2}$ = 2,454. Multigroup MRQAPs with 5,000 Y-permuted samples. * $p < .05$, ** $p < .01$, *** $p < .001$. We report p-values, because confidence intervals cannot be computed for MRQAPs. Various robustness analyses (the two samples separately, a standard linear regression, with a non-log-transformed dependent matrix, with non-merged RFID data, and including Big Five personality traits) are reported in Table S1 and Table S2 of the Supplementary Materials.

The results of the multi-group MRQAPs support the notion of *depression isolation*; dyads with a high mean in depressive symptoms were less likely to interact. It has to be noted that the effect size of the estimate cannot be interpreted directly due to the log transformation of the dependent matrix. The following example should illustrate the size of this effect: The interaction time between two individuals with a depression score of 5 each is estimated to be 9.12 seconds per hour (exp(2.504-0.059*5)), whereas an interaction between two individuals with a depression score of 20 is estimated to last for only 3.76 seconds per hour (exp(2.504-0.059*20); considering that everything else is the reference category - for instance, that there is no friendship tie present).

There was a positive effect for depression similarity; this suggests that social interactions were more likely between individuals that reported a similar level of depressive symptoms. Moreover, the interaction between depression mean and depression similarity was a negative predictor of social interactions, showing that depression homophily is stronger at the lower (the less depressed) end of the scale.

The multivariate interplay between predictors of social interactions and their effect size can be shown with a *selection table* where the estimates of a multivariate analysis are used to calculate an estimate for the dependent variable (i.e., social interaction duration) for various configurations of the predictors (e.g., see Elmer et al., 2017). In our case, we want to show how various levels of depressive symptoms of individual $i$ and $j$ predict social interaction duration of the dyad $y_{ij}$ with the estimates of depression mean, depression similarity, and their interaction. The values for $\widehat{y}_{ij}$ of the observed range of depressive symptoms of $i$ and $j$ (0 to 36) are shown in a heatmap in Figure 3. Details on the computation of $\widehat{y}_{ij}$ for this Figure can be found in the Supplementary Material (Section Computation of the Selection Table). For the case of two male students of sample one that are not friends and have the same age (i.e., all reference categories), Figure 3 shows that interactions where both individuals were highly depressed were the least likely and those most likely were interactions between low depressed individual or when one individual was highly depressed and the other one low in depression.

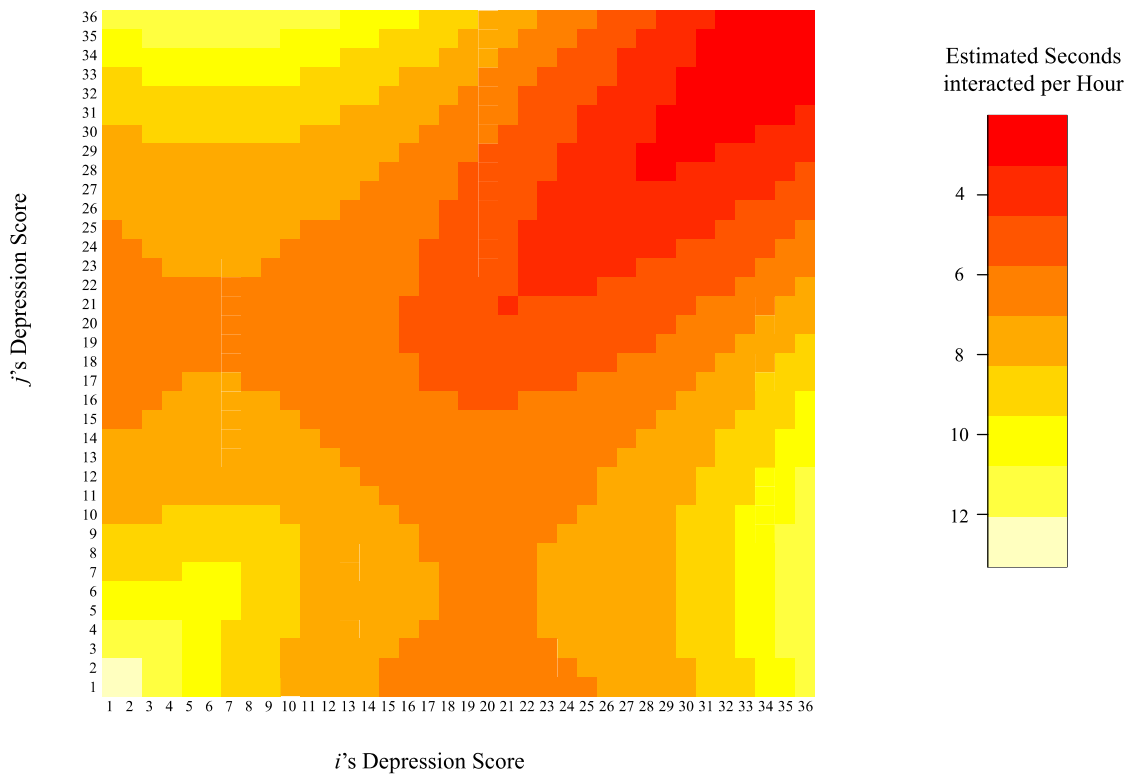

*Figure 3.* $\hat{y}_{ij}$ (in sec/h) for depression values between 1 and 36 (i.e., the range of observed values) for the case of all reference categories (of e.g., gender, age, friendship ties).

To investigate how depressive symptoms are associated with the extent to which individuals interact with friends, we tested an interaction of depression mean with the symmetrized friendship matrix. There was no significant effect of depression mean with being friends in predicting social interactions. As noted earlier, friendship relations might be mutual or asymmetric (either both individuals consider the relationship as a friendship or just one of the two). Neglecting this information might diffuse the differentiation between weak and strong friendship ties. For this reason, we conducted additional analyses in which we included two matrices capturing the mutual and asymmetric friendship relations instead of one symmetrized friendship matrix.

In those analyses we find a negative interaction effect of depression mean with being mutual friends in predicting social interactions ($\beta$ = -0.084, p = .029), indicating that depressed individuals tend to interact less with their reciprocated friends than non-depressed individuals. Interestingly, the interaction of asymmetric friendship ties with depression mean was positive but did not predict interaction duration significantly ($\beta$ = 0.069, p = .095). Details on these results are provided in Table S3 of the Supplementary Materials.

Beyond these depression-related findings, the multi-group MRQAP analysis shows significantly higher estimates for sample two and the estimates increased with an increasing mean age and increasing age similarity. Negative estimates were found for both individuals being female, indicating that interactions between two females are less observed than between two males.

We conducted a number of robustness analyses of these multi-group MRQAP analyses: (1) for the two samples separately, with a (2) non log-transformed dependent matrix, and (3) with non-merged RFID data (interactions of dyads that were no longer than 75 seconds apart, have been merged as recommended by Elmer et al. (2018) for an improved validity). Also, we included measures of the Big Five personality traits into the model. The results of these robustness analyses can be found in Table S1 and Table S2 of the Supplementary Material. All these robustness analyses yield that the finding of this study are robust against different data treatments, within each sample, and when controlling for the effect of personality traits. The exception being the depression similarity effect, which is not a significant predictor in the separate analysis of sample two ($\beta = 0.024$, p = .142) and when modeling the non log-transformed duration matrix ($\beta = 0.557$, p = .213), and the depression mean effect which is not significant when modeling the non log-transformed duration matrix ($\beta = -0.474$, p = .160).

**Dyadic and group interactions**

Finally, we tested the assumption that individuals with depressive symptoms spend relatively more time in dyadic interactions than in group interactions (*dyadic isolation*). This hypothesis cannot be tested with the MRQAP as the unit of analysis is beyond a dyadic relation. To account for the interdependencies between observations we performed a permutation-based correlation test of depressive symptoms on the ratio of dyadic interactions in all social interactions (i.e., $\frac{x_{dyad}}{x_{dyad}+x_{group}}$, where $x_i$ indicates the number of seconds spent in an interaction of type $i$). Permuting the dependent variable (i.e., the ratio) here follows the general logic of bivariate QAPs (Krackhardt, 1988). There was a positive correlation between an individual's ratio of dyadic interactions and depressive symptoms and ($r$(121) = .263, p = .003, 5.000 Y-permutations). In other words, the more depressive symptoms an individual reports, the smaller is the proportion of group interactions of the total time spent in social interactions.

**Discussion and conclusion**

In this study, we investigated how individuals' depressive symptoms affect social interactions within two independent student communities spending a weekend socializing in a remote camp house. We find that individuals' depressive symptoms are associated with spending less time in social interactions. This is in line with our first hypothesis (depression-isolation hypothesis). We also find that individuals tend to interact with others that have a similar level of depressive symptoms, as predicted by our second hypothesis (depression-homophily hypothesis). This homophily effect is more pronounced on the lower end of the depression scale. We find no support for the third hypothesis (depression-friendship hypothesis), stating that individuals' depressive symptoms are associated with the extent to which they interact with friends. In further explorations, we find that the likelihood of interacting with mutual friends (i.e., both individuals nominating each other) decreases with higher depression scores. We find no such effects for asymmetric friendship ties (i.e., only one friendship nomination). This might indicate that the hypothesized association depends of the strength of a friendship relation. In line the fourth hypothesis (dyadic-isolation hypothesis), depressive symptoms are associated with the sizes of interaction groups; individuals high in depressive symptoms are more likely to interact in pairs than in groups.

These findings contribute to the broad literature on the association between depressive symptoms and social interactions (e.g., Brown et al., 2011; Nezlek et al., 1994). Many prior studies rely on depression and interaction self-reports. However, social network research designs are necessary to explore more complex relational phenomena. Besides generally lower levels of social interactions, network-specific behavior patterns of depressed individuals can additionally contribute to their vicious cycle of social isolation and depression. First, the tendency to interact with similarly depressed individuals can lead to more exposure to their dysfunctional attitudes and thus being socially influenced to develop more depressive symptoms (van Zalk, Kerr, Branje, Stattin, & Meeus, 2010). Second, because of the unique support that strong friends can provide (e.g., emotional support) a lack of interactions with those can lead to the development of more symptomology (Lin, Ye, & Ensel, 1999). Third, the tendency of depressed individuals to interact in pairs instead of groups, could additionally contribute to the interaction partners' social isolation.

To study the network dimension of social interaction and depression, we apply established social network analysis methods (i.e., MRQAP; Dekker et al., 2007). These consider that observations were not sampled randomly from a large population (like in most other psychological studies) but consisted of a closed community of individuals where the dependence *between* individual's depressive symptoms was at the core of the analysis (e.g., how likely is an interaction based on the similarity in depressive symptoms of two individuals). MRQAPs follow the general estimation intuition of a multivariate regression and are thus straightforwardly interpreted as illustrated in our results.

The empirical setting of this study was unique in many ways. First, we measured social interactions with recently developed RFID badges that allowed us to observe individual behavior directly. Given the small number of studies on "actual" behavior, scholars have been encouraged to applying such methods to approach psychological research questions (Baumeister, Vohs, & Funder, 2007). Second, we combined these data with state of the art sociometric data (friendships) and self-report data of depressive symptoms.

Third, the fact that the students spent an entire weekend in a remote camp house, constituted an isolated setting where only social interaction between participants were possible. All attendees of the weekends participated in the RFID data collection, providing us with a full-range view on the social interaction dynamics of the participating individuals.

This study also had a number of limitations. First, our empirical setting was in a very specific population and context – a socializing weekend of first semester students. Presumably all participants felt a norm of being socially engaged at this event. At the same time, friendship relations were often formed relatively recently. Future studies should investigate social interaction networks in different social settings and aim at oversampling clinically depressed individuals. Second, our method of measuring social interaction was limited to assessing quantitative aspects of a social interaction but not qualitative aspects of the social interactions. Hence, we do not know how a potential social skill deficit of depressed individuals actually affected characteristics of social interactions (e.g., eye contact avoidance of depressed individuals; Segrin, 2000). Third, we aggregated the social interactions of the two samples for the time of the data-collection period and thus leave out the temporal dynamics of these social interactions. Depressive symptoms could be related to particular interaction sequences, for which time-stamped network analysis methods could be a suitable framework (Butts, 2008; Christoph Stadtfeld, Hollway, & Block, 2017). Fourth, the undirected nature of the social interaction measure only allows us to draw conclusions about which interactions are more likely—and not which interactions depressed individuals seek, avoid, or terminate.

Despite these limitations, our study has highlighted the strong effects that an individual's depressive symptoms have on social interactions. We have further demonstrated that social network designs and methodologies can offer us new insights on fundamental issues of psychology. We believe that an in-depth understanding of the small-scale social consequences of depressive symptoms can help to design interventions targeting the downward spiral of depression and social isolation more effectively.

**References**

Backenstrass, M., Frank, A., Joest, K., Hingmann, S., Mundt, C., & Kronm, K. (2006). A

comparative study of nonspecific depressive symptoms and minor depression regarding

functional impairment and associated characteristics in primary care. *Comprehensive*

*Psychiatry*, *47*, 35–41. http://doi.org/10.1016/j.comppsych.2005.04.007

Baddeley, J. L., Pennebaker, J. W., & Beevers, C. G. (2012). Everyday Social Behavior During a

Major Depressive Episode. *Social Psychological and Personality Science*, *4*(4), 445–452.

http://doi.org/10.1177/1948550612461654

Barnett, P. A., & Gotlib, I. H. (1988). Psychosocial functioning and depression: Distinguishing

among antecedents, concomitants, and consequences. *Psychological Bulletin*, *104*(1), 97–

126. http://doi.org/10.1037/0033-2909.104.1.97

Baumeister, R. F., Vohs, K. D., & Funder, D. C. (2007). Psychology as the Science of Self-

Reports and Finger Movements: Whatever Happened to Actual Behavior? *Perspectives on*

*Psychological Science*, *2*(4), 396–403. http://doi.org/10.1111/j.1745-6916.2007.00051.x

Berry, D. S., & Hansen, J. S. (1996). Positive affect, negative affect, and social interaction.

*Journal of Personality and Social Psychology*, *71*(4), 796–809. http://doi.org/10.1037/0022-

3514.71.4.796

Brandes, U., & Wagner, D. (2004). Analysis and visualization of social networks. *Graph*

*Drawing Software*, 1–20. http://doi.org/10.1007/3-540-45848-4_47

Brown, L. H., Strauman, T., Barrantes-Vidal, N., Silvia, P. J., & Kwapil, T. R. (2011). An

experience-sampling study of depressive symptoms and their social context. *The Journal of*

*Nervous and Mental Disease*, *199*(6), 403–409.

http://doi.org/10.1097/NMD.0b013e31821cd24b

Burnett Heyes, S., Jih, Y. R., Block, P., Hiu, C. F., Holmes, E. A., & Lau, J. Y. F. (2015).

Relationship Reciprocation Modulates Resource Allocation in Adolescent Social Networks:

Developmental Effects. *Child Development*, *86*(5), 1489–1506.

http://doi.org/10.1111/cdev.12396

Butts, C. T. (2008). A Relational Event Framework for Social Action. *Sociological Methodology*,

*38*(1), 155–200. http://doi.org/10.1111/j.1467-9531.2008.00203.x

Cattuto, C., van den Broeck, W., Barrat, A., Colizza, V., Pinton, J. F., & Vespignani, A. (2010).

Dynamics of person-to-person interactions from distributed RFID sensor networks. *PLoS

ONE*, *5*(7), 1–9. http://doi.org/10.1371/journal.pone.0011596

Coyne, J. C. (1976a). Depression and the response of others. *Journal of Abnormal Psychology*,

*85*(2), 186–193.

Coyne, J. C. (1976b). Towards an interactional description of depression. *Psychiatry:

Interpersonal and Biological Processes*, *39*(1), 28–40.

http://doi.org/10.1521/00332747.1976.11023874

Dekker, D., Krackhardt, D., & Snijders, T. A. B. (2007). Sensitivity of MRQAP tests to collinearity and autocorrelation conditions. *Psychometrika*, *72*(4), 563–581.

http://doi.org/10.1007/s11336-007-9016-1

Elmer, T., Boda, Z., & Stadtfeld, C. (2017). The co-evolution of emotional well-being with weak and strong friendship ties. *Network Science*, *5*(3), 278–307.

http://doi.org/https://doi.org/10.1017/nws.2017.20

Elmer, T., Chaitanya, K., Purwar, P., & Stadtfeld, C. (2018). The validity of RFID badges
measuring face-to-face interactions. *Under review for publication*.

Friedkin, N. E. (1990). A Guttman Scale for the Strength of an Interpersonal Tie. *Social
Networks*, *12*, 239–252.

Gadassi, R., & Rafaeli, E. (2015). Interpersonal perception as a mediator of the depression-
interpersonal difficulties link: A review. *Personality and Individual Differences*, *87*, 1–7.
http://doi.org/10.1016/j.paid.2015.07.023

Gotlib, I. H. (1983). Perception and recall of interpersonal feedback: Negative bias in depression.
*Cognitive Therapy and Research*, *7*(5), 399–412. http://doi.org/10.1007/BF01187168

Gotlib, I. H., Lewinsohn, P. M., & Seeley, J. R. (1995). Symptoms versus a diagnosis of
depression: differences in psychosocial functioning. *Journal of Consulting and Clinical
Psychology*, *63*(1), 90–100. http://doi.org/10.1037/0022-006X.63.1.90

Hautzinger, M., & Bailer, M. (1993). *Allgemeine Depressionsskala [General Depression Scale]*.
Göttingen, Germany: Hogrefe Verlag.

Holt-Lunstad, J., Smith, T. B., Baker, M., Harris, T., & Stephenson, D. (2015). Loneliness and
Social Isolation as Risk Factors for Mortality: A Meta-Analytic Review. *Perspectives on
Psychological Science*, *10*(2), 227–237. http://doi.org/10.1177/1745691614568352

Jose, P. E., & Lim, B. T. L. (2014). Social connectedness predicts lower loneliness and
depressive symptoms over time in adolescents. *Open Journal of Depression*, *03*(04), 154–
163. http://doi.org/10.4236/ojd.2014.34019

Kawachi, I., & Berkman, L. F. (2001). Social ties and mental health. *Journal of Urban Health*, *78*(3), 458–467. http://doi.org/10.1093/jurban/78.3.458

Krackhardt, D. (1988). Predicting with networks: Nonparametric multiple regression analysis of dyadic data. *Social Networks*, *10*(4), 359–381. http://doi.org/10.1016/0378-8733(88)90004-4

Lewinsohn, P. M. (1974). A behavioral approach to depression. In *The psychology of depression: Contemporary theory and research* (pp. 157–78).

Libet, J. M., & Lewinsohn, P. M. (1973). Concept of Social Skill With Special Reference To the Behavior of Depressed Persons. *Journal of Consulting and Clinical Psychology*, *40*(2), 304–312. http://doi.org/10.1037/h0034530

Lin, N., Ye, X., & Ensel, W. M. (1999). Social support and depressed mood: A structural analysis. *Journal of Health and Social Behavior*, *40*(4), 344–359. http://doi.org/10.2307/2676330

McPherson, M., Smith-Lovin, L., & Cook, J. M. (2001). Birds of a feather: Homophily in social networks. *Annual Review of Sociology*, *27*, 415–444. http://doi.org/10.1146/annurev.soc.27.1.415

Nezlek, J. B., Hampton, C. P., & Shean, G. D. (2000). Clinical depression and day-to-day social interaction in a community sample. *Journal of Abnormal Psychology*, *109*(1), 11–9. http://doi.org/10.1037//0021-843X.109.1.11

Nezlek, J. B., Imbrie, M., & Shean, G. D. (1994). Depression and everyday social interaction. *Journal of Personality and Social Psychology*, *67*(6), 1101–11.

Radloff, L. S. (1977). The CES-D Scale: A Self-Report Depression Scale for Research in the General Population. *Applied Psychological Measurement*, *1*(3), 385–401. http://doi.org/10.1177/014662167700100306

Rook, K. S., Pietromonaco, P. R., & Lewis, M. a. (1994). When are dysphoric individuals distressing to others and vice versa? Effects of friendship, similarity, and interaction task. *Journal of Personality and Social Psychology*, *67*(3), 548–59. http://doi.org/10.1037/0022-3514.67.3.548

Rose, A. J. (2002). Co-rumination in the friendships of girls and boys. *Child Development*, *73*(6), 1830–1843. http://doi.org/10.1111/1467-8624.00509

Schaefer, D. R., Kornienko, O., & Fox, A. M. (2011). Misery does not love company: Network selection mechanisms and depression homophily. *American Sociological Review*, *76*(5), 764–785. http://doi.org/10.1177/0003122411420813

Segrin, C. (2000). Social skills deficits associated with depression. *Clinical Psychology Review*, *20*(3), 379–403. http://doi.org/10.1016/S0272-7358(98)00104-4

Stadtfeld, C., Hollway, J., & Block, P. (2017). Dynamic Network Actor Models: Investigating Coordination Ties through Time. *Sociological Methodology*, 008117501770929. http://doi.org/10.1177/0081175017709295

Stadtfeld, C., Vörös, A., Elmer, T., Boda, Z., & Raabe, I. (in press). Falling through the network: Social networks explain academic failure and success. *Proceedings of the National Academy of Sciences*.

Steptoe, A., Shankar, A., Demakakos, P., & Wardle, J. (2013). Social isolation, loneliness, and all-cause mortality in older men and women. *Proceedings of the National Academy of Sciences*, *110*(15), 5797–5801. http://doi.org/10.1073/pnas.1219686110

Valtorta, N. K., Kanaan, M., Gilbody, S., Ronzi, S., & Hanratty, B. (2016). Loneliness and social isolation as risk factors for coronary heart disease and stroke: Systematic review and meta-analysis of longitudinal observational studies. *Heart*, *102*(13), 1009–1016. http://doi.org/10.1136/heartjnl-2015-308790

van Zalk, M. H. W., Kerr, M., Branje, S. J. T., Stattin, H., & Meeus, W. H. J. (2010). Peer contagion and adolescent depression: The role of failure anticipation. *Journal of Clinical Child and Adolescent Psychology*, *39*(6), 837–848. http://doi.org/10.1080/15374416.2010.517164

**Supplementary Material:**

**Social interaction networks and depressive symptoms**

## Robustness Analyses

In the following section, we have conducted additional analyses to test the robustness of our findings. For this, we have run (1) a regular linear regression model (Table S1) and MRQAPs with (2) a non-log-transformed dependent variable (Table S1), (3) non-merged RFID data (interactions of dyads that were no longer than 75 seconds apart, have been merged as recommended by Elmer et al. (2018) for an improved validity; Table S1), (4) the two samples separately (Table S1), and (5) including measures of the Big Five personality traits (Table S2). All these analyses confirm that the main findings of the article are robust against different ways of treating the data. Only the depression similarity effect did not replicate in sample two and with a non-log-transformed dependent variable, as well as the depression isolation effect with a non-log-transformed dependent variable.

In the preregistration of this study[2], we intended to investigate not only the effect of depressive symptoms on the interactions with friends, but also with acquaintances. We omitted the acquaintance network from this analysis to reduce the complexity of the story of the paper. Nevertheless, for an additional analysis we substituted the friendship network with the acquaintance network, which contained all friendship nominations. The main findings of the study remain robust and the interactions of the depression mean matrix with the acquaintance network show a similar pattern ($\beta_{\text{mutual acquaintance * depression mean}} = -0.118$, p = .007, $\beta_{\text{asymmetric acquaintance * depression mean}} = 0.022$, p = .720).

---

[2] www.osf.io/xce9g

Moreover, the preregistered analysis plan was modified in two ways: First, testing the depression isolation hypothesis with a matrix, in which all cells in a row contain the depression value of the individual, was not feasible due to the undirected nature of the social interaction network. Hence, we had to aggregate the depression scores to the dyadic level and take the mean depression value. Second, the dependent variable was log-transformed because the linear regression assumption of normality of errors was violated when modeling the non-log-transformed dependent variable.

Table S1

*Results of five models using different data sources or statistical inference methods*

| | Linear Regression | | | Non-Transformed | | Non-Merged | | Sample 1 | | Sample 2 | |
| | | 95% CI | | | | | | | | | |
| | Est. | lower | upper | Est. | p | Est. | p | Est. | p | Est. | p |
|---|---|---|---|---|---|---|---|---|---|---|---|
| intercept | 2.504*** | 1.598 | 2.534 | -9.630* | .018 | 2.066* | .022 | 2.926*** | <.001 | 3.163† | .062 |
| sample two | 0.806*** | 0.532 | 0.842 | 10.759 | .141 | 0.687 | .288 | - | - | | |
| at least one female | -0.095 | -0.239 | 0.040 | -1.386 | .312 | -0.100† | .078 | -0.194* | .024 | 0.024 | .435 |
| both female | -0.148 | -0.268 | 0.018 | 0.660 | .423 | -0.125* | .046 | 0.02 | .433 | -0.278** | .003 |
| age mean (centered) | 0.065* | 0.008 | 0.108 | 0.259 | .395 | 0.058* | .011 | 0.008 | .445 | 0.116** | .002 |
| age similarity | 0.042* | 0.011 | 0.073 | 1.200* | .046 | 0.042** | .007 | 0.033 | .124 | 0.050* | .016 |
| one student organization[1] | -0.028 | -0.315 | 0.477 | 32.489** | .004 | 0.081 | .338 | - | - | -0.189 | .211 |
| same student status[1] | 0.269 | -0.059 | 0.719 | 38.114** | .001 | 0.330† | .058 | - | - | 0.209 | .176 |
| being friends | 2.128*** | 1.628 | 2.445 | 104.378*** | <.001 | 2.036*** | <.001 | 2.265*** | <.001 | 1.974*** | <.001 |
| depression mean | -0.059*** | -0.068 | -0.027 | -0.474 | .160 | -0.047*** | <.001 | -0.064*** | .001 | -0.042** | .003 |
| depression similarity | 0.047** | 0.008 | 0.067 | 0.557 | .213 | 0.038** | .008 | 0.069* | .018 | 0.024 | .142 |
| depression mean * depression similarity | -0.004*** | -0.005 | -0.001 | -0.016 | .344 | -0.003** | .002 | -0.004* | .041 | -0.002* | .036 |
| depression mean * being friends | -0.012 | -0.052 | 0.041 | 1.958† | .053 | -0.006 | .404 | 0.036 | .270 | -0.013 | .349 |
| $R^2$ | .123 | | | .168 | | .128 | | .068 | | .125 | |
| Adj. $R^2$ | .119 | | | .164 | | .124 | | .062 | | .116 | |

*Note.* $N_{sample1+2}$ = 2,454, (Multigroup) MRQAPs with 5,000 Y-permuted samples. CI = confidence interval, † $p < .10$, * $p < .05$, ** $p < .01$, *** $p < .001$.

Non-Merged = the interactions were not merged according to the recommendations of (Elmer et al., 2018), Non-Transformed = the dependent variable (interaction durations) has not been log-transformed. [1] no estimate of sample one because of missing depression scores of student organization members.

**Multi-group MRQAP controlling for personality traits**

To take into account that an individual's personality can affect the formation of social interactions, we conducted a further multi-group MRQAP analysis including effects for the Big Five personality traits. The Big Five personality traits were assessed with the 10 item version of the Big Five Inventory (BFI; Rammstedt & John, 2007) where every trait is measured with two items each, rated on a 5-point Likert scale ranging from "disagree strongly" (1) to "agree strongly" (5). A sample item for neuroticism is "I see myself as someone who: is relaxed, handles stress well" (inverse coded). The internal consistency of these item varied from $\alpha = .33$ (agreeableness) to $\alpha = .80$ (extraversion) and the mean values were between 2.87 (neuroticism) and 3.61 (openness). We constructed two matrices for each trait that represent the centered mean value of $i$ and $j$ in the respective trait as well as their similarity in that trait. The results of this multi-group MRQAP analysis including personality traits can be found in Table S2. The findings of hypothesis one and two were robust in size and significance when including personality traits in the model.

Table S2

*Multi-group MRQAP results on log transformed interaction durations of dyads including*

*personality traits*

|  | Est. | p |
|---|---|---|
| intercept | 3.363*** | <.001 |
| sample two | 0.774 | .242 |
| at least one female | -0.431*** | <.001 |
| both female | -0.221** | .008 |
| age mean (centered) | 0.078** | .006 |
| age similarity | -0.041* | .016 |
| one student organization | -0.326† | .080 |
| same student status | 0.132 | .278 |
| being friends | 2.079*** | <.001 |
| openness mean (centered) | -0.135* | .015 |
| openness similarity | 0.087* | .028 |
| conscientiousness mean (centered) | 0.369*** | <.001 |
| conscientiousness similarity | -0.046 | .180 |
| extraversion mean (centered) | -0.126* | .017 |
| extraversion similarity | 0.066† | .076 |
| agreeableness mean (centered) | -0.278*** | <.001 |
| agreeableness similarity | -0.106* | .026 |
| neuroticism mean (centered) | 0.202*** | <.001 |
| neuroticism similarity | 0.001 | .477 |
| depression mean | -0.081*** | <.001 |
| depression similarity | 0.051** | .003 |
| depression mean * depression similarity | -0.004*** | <.001 |
| depression mean * being friends | -0.010 | .362 |
| $R^2$ | .143 | |
| Adj. $R^2$ | .134 | |

*Note.* $N = 2,454$, Multigroup MRQAP with 5,000 Y-permuted samples. † $p < .10$, * $p < .05$, ** $p < .01$, *** $p < .001$.

**The Role of Friendship Strength**

To explore the role of friendship strength when testing the *depression-friendship hypothesis* we conduced additional analyses in which we differentiate between asymmetric and mutual friendship ties. We generally assume that mutual friendship ties (i.e., person A is nominating person B and person B is nominating person A) are stronger than asymmetric friendship ties (i.e., only either person A nominates person B or person B nominates person A; Friedkin, 1990). Table S3 shows the results of the MRQAP model including separate effects for mutual and asymmetric friendship ties. These results indicate that depressive symptoms are associated with spending less time with strong (i.e., mutual) friends. We find no effect of the interaction of asymmetric friendship ties with the depression mean matrix when predicting social interactions. When conducing the above-mentioned robustness checks (see Section Robustness Analyses) with this model specification, we consistently find support that depressive symptoms are associated with interacting less with strong friends (model results are not reported here due to space restrictions). Also, the findings of the depression-isolation and the depression-homophily hypotheses remain robust with this alternative model specification.

Table S3

*Multi-group MRQAP results on log transformed interaction durations of dyads*

|  | Est. | p |
|---|---|---|
| intercept | 2.501** | .006 |
| sample two | 0.794 | .268 |
| at least one female | -0.099 | .106 |
| both female | -0.152* | .038 |
| age mean (centered) | 0.062* | .019 |
| age similarity | 0.041* | .012 |
| one student organization | -0.012 | .487 |
| same student status | 0.276 | .108 |
| being mutual friends | 3.102*** | <.001 |
| being asymmetric friends | 1.224* | .019 |
| depression mean | -0.059*** | <.001 |
| depression similarity | 0.047** | .002 |
| depression mean * depression similarity | -0.004*** | <.001 |
| depression mean * being mutual friends | -0.084* | .029 |
| depression mean * being asymmetric friends | 0.069 | .095 |
| $R^2$ | .125 | |
| Adj. $R^2$ | .120 | |

*Note.* $N_{sample1+2} = 2,454$, Multigroup MRQAP with 5,000 Y-permuted samples. † $p < .10$, * $p < .05$, ** $p < .01$, *** $p < .001$.

**Computation of the Selection Table**

In Figure 3 of the main text we show the likelihood of interactions between two individuals $i$ and $j$ based on their depressive symptoms (i.e., $v_i$ and $v_j$). Here we report how we computed these values.

Considering that everything else is the reference category (e.g., male dyad interaction, non-friends), we get the following formula for the estimate of interaction duration of the dyad $\widehat{y}_{ij}$ based on depression mean and depression similarity:

$$\widehat{y}_{ij} = \exp((\beta_0 + \beta_1 \frac{v_i + v_j}{2} + \beta_2(-1)|v_i - v_j| + \beta_3 \frac{v_i + v_j}{2}(-1)|v_i - v_j|)$$

where $\beta_{0,1,2,3}$ denote the estimates for the intercept, depression mean, depression similarity and the interaction of the depression mean and the depression similarity matrix. Variables $v_i$ and $v_j$ are the depression scores of actor $i$ and $j$. Taking the respective estimates from Table 2 of the main text for values of the observed depression scores we obtained a table of estimated $\widehat{y}_{ij}$. If we consider two example dyads, one where individuals report a depression score of $v_i = 4$ and $v_j = 6$, and the other where values of $v_i = 4$ and $v_k = 16$ were reported, we would get estimates of $\widehat{y}_{ij} = 8.61$ seconds per hour and $\widehat{y}_{ik} = 6.00$ seconds per hour, respectively. Using the above introduced equation we then computed $\widehat{y}_{ij}$ for each combination of the observerd values of depressive symptoms to construct the table/heatmap reported Figure 3 of the main text.

**References**

Elmer, T., Chaitanya, K., Purwar, P., & Stadtfeld, C. (2018). The validity of RFID badges measuring face-to-face interactions. *Under review for publication.*

Friedkin, N. E. (1990). A Guttman Scale for the Strength of an Interpersonal Tie. *Social Networks*, *12*, 239–252.

Rammstedt, B., & John, O. P. (2007). Measuring personality in one minute or less: A 10-item short version of the Big Five Inventory in English and German. *Journal of Research in Personality*, *41*(1), 203–212. http://doi.org/10.1016/j.jrp.2006.02.001